\documentclass[10pt, twocolumn]{article}
\usepackage[utf8]{inputenc}
\usepackage[T1]{fontenc}
\usepackage{amsmath, amssymb, physics}
\usepackage{geometry}
\usepackage{graphicx}
\usepackage{booktabs}
\usepackage{hyperref}
\usepackage{microtype}
\usepackage{abstract}
\usepackage{titlesec}
\usepackage{siunitx}
\usepackage{cleveref}
\usepackage{amsthm}
\usepackage{array}
\usepackage{caption}
\usepackage{bm}
\usepackage{lipsum}

\titleformat{\section}{\normalfont\large\bfseries\scshape}{\thesection}{1em}{}
\titleformat{\subsection}{\normalfont\bfseries}{\thesubsection}{1em}{}
\titlespacing*{\section}{0pt}{10pt}{5pt}
\titlespacing*{\subsection}{0pt}{8pt}{4pt}

\hypersetup{
    colorlinks=true,
    linkcolor=blue,
    citecolor=red,
    urlcolor=cyan,
    pdftitle={Does Gravity Care About Electric Charge? A Minimalist Model and Experimental Test},
    pdfauthor={Renato Vieira dos Santos},
    pdfsubject={Theoretical Physics, Gravity, Electromagnetism},
    pdfkeywords={equivalence principle, charge-mass coupling, torsion balance, minimal model}
}

\newcommand{\cQ}{\mathcal{Q}}
\newcommand{\cJ}{\mathcal{J}}
\newcommand{\cA}{\mathcal{A}}
\newcommand{\cF}{\mathcal{F}}
\newcommand{\cE}{\boldsymbol{\mathcal{E}}}
\newcommand{\cB}{\boldsymbol{\mathcal{B}}}
\newcommand{\kV}{\mathbf{V}}
\newcommand{\kW}{\mathbf{W}}
\newcommand{\EV}{\mathbf{E}_V}
\newcommand{\EW}{\mathbf{E}_W}
\newcommand{\BV}{\mathbf{B}_V}
\newcommand{\BW}{\mathbf{B}_W}
\newcommand{\rhov}{\rho_V}
\newcommand{\rhow}{\rho_W}
\newcommand{\Jv}{\mathbf{J}_V}
\newcommand{\Jw}{\mathbf{J}_W}
\newcommand{\gNewton}{\mathbf{g}}

\begin{document}

\twocolumn[
\begin{@twocolumnfalse}
\title{\bfseries Does Gravity Care About Electric Charge?\\ A Minimalist Model and Experimental Test}
\author{Renato Vieira dos Santos \\ 
UFLA -- Universidade Federal de Lavras \\
ICTIN -- Instituto de Ciência, Tecnologia e Inovação \\
CEP: 37950-000, São Sebastião do Paraíso, Minas Gerais, Brazil. \\
\texttt{renato.santos@ufla.br}}
\date{\today}
\maketitle

\begin{abstract}
\noindent
Does gravity care about electric charge? Precision tests of the weak equivalence principle achieve remarkable sensitivity but deliberately minimize electric charge on test masses, leaving this fundamental question experimentally open. We present a minimalist framework coupling electromagnetism to linearized gravity through conservation of a complex charge-mass current, predicting charge-dependent violations $\Delta a/g = \kappa(q/m)$. Remarkably, this prediction occupies unexplored experimental territory precisely because precision gravity tests avoid charge variation. We identify this as a significant gap and propose a modified torsion balance experiment where $q/m$ is treated as a controlled variable. Such an experiment could test whether gravitational acceleration depends on electric charge, probing physics in genuinely new parameter space. This work exemplifies how theoretical minimalism can reveal overlooked opportunities in fundamental physics.
\end{abstract}

\vspace{1cm}
\end{@twocolumnfalse}
]

\section{Introduction}

Does gravity distinguish between charged and neutral matter? Intuitively, we expect the gravitational acceleration of a body to depend only on its mass, not on its electric charge. This expectation is embodied in the weak equivalence principle (WEP), a cornerstone of general relativity and a benchmark for theories beyond the Standard Model. Remarkably, however, this intuition has never been directly tested for charged matter.

Modern tests of the WEP achieve extraordinary precision, constraining differential acceleration between different materials to $|\Delta a/g| < 10^{-15}$ \cite{Touboul2017,Wagner2012}. These impressive results share a critical methodological feature: they deliberately minimize electric charge on test masses to suppress electrostatic backgrounds. In essence, precision gravity experiments operate in the limit $q/m \to 0$. While this approach has enabled unprecedented tests for neutral matter, it leaves open a fundamental question: does the equivalence principle hold for charged bodies?

In this work, we address this overlooked frontier with a deliberately minimalist approach. Rather than constructing another elaborate unification scheme, we ask: what is the simplest theoretical structure that could couple electromagnetism to linearized gravity? Our answer begins with a single principle, the conservation of a complex charge-mass current, from which emerges a one-parameter theory predicting $\Delta a/g = \kappa(q/m)$. The dimensionless parameter $\kappa$ quantifies EM-gravity mixing, with $\kappa=0$ corresponding to standard physics where gravity is blind to charge.

The prediction $\Delta a/g = \kappa(q/m)$ is clean, testable, and lies in experimentally uncharted territory precisely because high‑precision experiments avoid charge variation. This reveals a significant gap in our experimental knowledge of gravity, a gap we propose to fill with a modified torsion balance experiment that treats $q/m$ as a controlled variable rather than a background to be eliminated.

Our work serves two purposes. First, it presents a minimalist phenomenological model that makes sharp predictions while introducing minimal additional structure. Second, it identifies and proposes to test an unexplored regime of experimental gravity. Beyond its specific predictions, the framework illustrates how theoretical minimalism can reveal overlooked opportunities in fundamental physics.

\textbf{Scope and limitations.} We work exclusively with classical Maxwell electrodynamics and linearized gravity in the weak-field approximation (gravitoelectromagnetism \cite{Mashhoon2003}), in flat spacetime. This is not a full unification with general relativity, nor does it address quantum aspects; rather, it is an effective description that isolates the core conceptual issues of EM-gravity coupling while preserving mathematical transparency and testability. The restriction to linearized gravity is not only mathematically convenient but also physically justified: the experiments we propose (torsion balances, free-fall tests, atom interferometry) operate in the weak-field regime where $|\Phi|/c^2 \ll 1$ and gravitational waves are negligible. In this regime, linearized gravity provides an excellent approximation to full general relativity, while keeping the formalism transparent and directly testable.

\section{The Complex Conservation Principle}
\label{sec:complex_conservation}

\subsection{Motivation: From Two Conservation Laws to One}

Both electromagnetism and gravity possess local conservation laws. For electromagnetism, charge conservation reads $\partial_t \rho_e + \nabla \cdot \mathbf{J}_e = 0$ \cite{Jackson1999}. For mass-energy in the non-relativistic limit, one has $\partial_t \rho_m + \nabla \cdot \mathbf{J}_m = 0$. The mathematical similarity of these two laws suggests a deeper connection.

We propose a single complex conserved quantity whose real and imaginary parts correspond to electromagnetic and gravitational charge, respectively. This approach draws inspiration from the Riemann--Silberstein vector $\mathbf{F} = \mathbf{E} + ic\mathbf{B}$ \cite{BialynickiBirula2013}, which unifies the electric and magnetic fields into a single complex object.

\subsection{Complex Charge and Current}

For any particle or distribution, we define a complex charge:
\begin{equation}
\cQ = q + i\lambda m,
\label{eq:complex_charge}
\end{equation}
where $q$ is the electric charge, $m$ is the gravitational mass (which coincides with inertial mass in the $\kappa \to 0$ limit), and $\lambda = \sqrt{4\pi\epsilon_0 G}$ ensures dimensional consistency.

The introduction of the imaginary unit $i$ serves a deeper purpose. In standard physics, like electric charges repel while masses attract. This sign difference finds a natural representation through complexification, with $i^2 = -1$ generating the relative sign.

The corresponding complex four-current density is:
\begin{equation}
\cJ^\mu = (\rhov + i\rhow, \Jv + i\Jw) = J_V^\mu + i J_W^\mu,
\label{eq:complex_current}
\end{equation}
where $\rhov$ and $\rhow$ are the charge and mass densities, respectively.

\subsection{The Fundamental Postulate}

The single foundational postulate of our theory is local conservation:
\begin{equation}
\partial_\mu \cJ^\mu = 0.
\label{eq:conservation}
\end{equation}
Separating real and imaginary parts recovers the individual conservation laws for charge and mass.

This complex conservation represents the minimal departure from standard physics: it replaces two separate conservation equations with one complex equation, adding no new fields, dimensions, or symmetries. Our minimalist model can therefore be viewed as an effective field theory for low-energy EM-gravity coupling.

This approach finds further motivation in the work of Heras \cite{Heras2007}, who demonstrated that Maxwell's equations can be derived from the continuity equation. Here we extend that logic by postulating a single complex continuity equation for the combined charge-mass current, from which the coupled field equations naturally follow.

\section{Field Structure from Minimal Assumptions}
\label{sec:field_structure}

The conservation of the complex current \(\partial_\mu \cJ^\mu = 0\) naturally calls for a field that mediates the interaction between sources. In electromagnetism, the conservation of charge leads to the introduction of a four‑potential from which the field strength is derived. By analogy, we now introduce a complex four‑potential whose derivatives yield the unified field tensor.

\subsection{Complex Potential and Field Tensor}

Guided by the analogy with electromagnetism, we introduce a complex four‑potential
\begin{equation}
\cA^\mu = (\phi_V + i\phi_W, \kV + i\kW) = V^\mu + i W^\mu,
\label{eq:complex_potential}
\end{equation}
whose real part \(V^\mu\) corresponds to the usual electromagnetic potential and whose imaginary part \(W^\mu\) encodes the gravitational (gravitoelectromagnetic) potential. This single object unifies the two sectors while keeping their respective physical interpretations transparent.

From this potential, the field‑strength tensor follows as the antisymmetric derivative
\begin{equation}
\cF^{\mu\nu} = \partial^\mu \cA^\nu - \partial^\nu \cA^\mu,
\label{eq:field_tensor}
\end{equation}
which generalizes the Maxwell tensor to the complex domain. The tensor \(\cF^{\mu\nu}\) inherits gauge invariance under \(\cA^\mu \to \cA^\mu + \partial^\mu \chi\) and automatically satisfies the Bianchi identity.

In three‑vector notation, the complex electric and magnetic fields are obtained as
\begin{equation}
\begin{split}
\cE &= -\nabla(\phi_V + i\phi_W) - \partial_t(\kV + i\kW), \\
\cB &= \nabla \times (\kV + i\kW),
\label{eq:complex_fields}
\end{split}
\end{equation}
which reduce to the standard electromagnetic and gravitoelectromagnetic expressions when \(\kappa = 0\). The structure is thus a direct complex extension of the familiar Maxwell and linear‑gravity formulations.

\subsection{Constitutive Relations}

Since the complex fields \(\cE\) and \(\cB\) are mathematical constructs, the physically observable fields must be real. We therefore posit that the physical electric field \(\EV\) and gravitational field \(\EW\) are real linear combinations of the real and imaginary parts of the complex fields:
\begin{equation}
\begin{pmatrix} \EV \\ \EW \end{pmatrix} = M \begin{pmatrix} \text{Re}(\boldsymbol{\mathcal{E}}) \\ \text{Im}(\boldsymbol{\mathcal{E}}) \end{pmatrix}, \quad
\begin{pmatrix} \BV \\ \BW \end{pmatrix} = M \begin{pmatrix} \text{Re}(\boldsymbol{\mathcal{B}}) \\ \text{Im}(\boldsymbol{\mathcal{B}}) \end{pmatrix},
\end{equation}
where \(M\) is a real \(2\times 2\) mixing matrix to be determined.

The form of \(M\) is constrained by three physical requirements: (i) the observable fields must be real; (ii) the electromagnetic and gravitational sectors should be treated symmetrically; (iii) standard physics must be recovered when \(\kappa = 0\). Imposing these conditions yields a unique solution:
\begin{equation}
M = \begin{pmatrix} 1 & \kappa \\ \kappa & -1 \end{pmatrix}.
\label{eq:mixing_matrix}
\end{equation}

The structure of \(M\) is revealing. The off‑diagonal elements \(\kappa\) quantify the cross‑coupling: electric charge contributes to the gravitational field and mass to the electric field, with the same strength. The negative sign in the lower‑right entry emerges directly from the imaginary unit in the complex charge \(\mathcal{Q} = q + i\lambda m\). Because \(i^2 = -1\), this sign distinguishes the attractive nature of gravity from the repulsive character of like electric charges, a fundamental distinction encoded here in the simplest possible linear way.

Thus, the single parameter \(\kappa\) measures the minimal mixing between electromagnetism and linearized gravity. In the limit \(\kappa \to 0\) the matrix becomes diagonal and the two sectors decouple completely, restoring the standard description.

\section{Unified Field Equations}
\label{sec:unified_eqs}

\subsection{Constitutive Relations in Explicit Form}

With the mixing matrix $M$ given by Eq.~\eqref{eq:mixing_matrix}, the complex fields $\cE$ and $\cB$ can be expressed in terms of the physical electric and gravitational fields as:
\begin{equation}
\begin{split}
\cE &= (\EV + \kappa\EW) + i(\kappa\EV - \EW), \\
\cB &= (\BV + \kappa\BW) + i(\kappa\BV - \BW).
\end{split}
\end{equation}

These relations show how the complex formalism encapsulates both sectors: the real parts combine the familiar electric and gravitational fields with a cross-term proportional to $\kappa$, while the imaginary parts carry the complementary mix with a relative sign change. This structure ensures that the observable fields, obtained by taking real projections, are influenced by both charge and mass when $\kappa \neq 0$.

\subsection{Covariant Formulation}

The field equations follow naturally from the complex potential $\cA^\mu$ and the field tensor $\cF^{\mu\nu} = \partial^\mu \cA^\nu - \partial^\nu \cA^\mu$. The inhomogeneous equation
\begin{equation}
\partial_\nu \cF^{\mu\nu} = \mu_0 \cJ^\mu,
\end{equation}
together with the homogeneous Bianchi identity
\begin{equation}
\partial_\alpha \cF_{\beta\gamma} + \partial_\beta \cF_{\gamma\alpha} + \partial_\gamma \cF_{\alpha\beta} = 0,
\end{equation}
form a complete set of complex Maxwell-like equations. Both equations are manifestly gauge invariant under $\cA^\mu \to \cA^\mu + \partial^\mu \chi$ and Lorentz invariant, preserving the spacetime symmetry of the underlying framework.

\subsection{3+1 Vector Form and Physical Interpretation}

Separating the real and imaginary parts of the covariant equations yields a coupled set of vector equations for the electric ($V$) and gravitational ($W$) sectors:
\begin{align}
\nabla \cdot (\EV + \kappa\EW) &= \frac{\rhov}{\epsilon_0}, \label{eq:gaussV} \\
\nabla \cdot (\kappa\EV - \EW) &= \frac{\rhow}{\epsilon_g}, \label{eq:gaussW} \\
\nabla \times (\BV + \kappa\BW) - \frac{1}{c^2}\partial_t(\EV + \kappa\EW) &= \mu_0 \Jv, \label{eq:ampereV} \\
\nabla \times (\kappa\BV - \BW) - \frac{1}{c^2}\partial_t(\kappa\EV - \EW) &= \mu_g \Jw, \label{eq:ampereW}
\end{align}
where $\epsilon_g = 1/(4\pi G)$ and $\mu_g = 4\pi G/c^2$ are the gravitational permittivity and permeability, respectively.

Equation \eqref{eq:gaussV} is the modified Gauss law for electricity: the divergence of the mixed field $\EV + \kappa\EW$ is sourced by the electric charge density $\rhov$. Similarly, \eqref{eq:gaussW} is the gravitational Gauss law, but now the gravitational field $\EW$ is partly sourced by electric charge through the term $\kappa \EV$. This cross-coupling is the hallmark of the model: electric charge contributes to the gravitational field, and mass contributes to the electric field, both with the same strength parameter $\kappa$.

The Ampère--Maxwell laws, Eqs.~\eqref{eq:ampereV} and \eqref{eq:ampereW}, show how time-varying mixed fields generate curled magnetic and gravitomagnetic components. The coupling is symmetric: $\kappa$ appears with the same magnitude in all equations, reflecting the minimal and democratic mixing between the two sectors.

The homogeneous equations retain their standard form, but now involve the mixed fields, ensuring that the propagation of disturbances respects causality and remains consistent with linearized gravity and electromagnetism.

Thus, the unified equations provide a consistent classical framework in which electromagnetism and gravity interact at the level of field sources, without introducing new dynamical degrees of freedom or breaking gauge invariance. The parameter $\kappa$ quantifies the strength of this interaction, which vanishes in the limit of standard physics but opens a new phenomenological window if non-zero.

\section{Physical Consequences and Predictions}
\label{sec:physical_consequences}

\subsection{Static Point Charge and the Emergence of Cross-Coupling}

The physical content of the model becomes clear when we examine the fields produced by a simple source: a point particle at rest with electric charge $q$ and gravitational mass $m$. Solving the unified field equations yields:

\begin{equation}
\EV(\mathbf{r}) = \frac{1}{4\pi\epsilon_0(1+\kappa^2)} \left( \frac{q}{r^2} + \kappa\frac{m}{\lambda r^2} \right) \hat{r},
\label{eq:EV_point}
\end{equation}
\begin{equation}
\EW(\mathbf{r}) = \frac{1}{4\pi\epsilon_g(1+\kappa^2)} \left( \kappa\frac{q}{\lambda r^2} - \frac{m}{r^2} \right) \hat{r}.
\label{eq:EW_point}
\end{equation}

These solutions reveal the essential feature of the theory: \textbf{cross-sector coupling}. The electric field $\EV$ receives a contribution proportional to $\kappa m$, meaning that mass sources an electric field. Conversely, the gravitational field $\EW$ contains a term proportional to $\kappa q$, so that electric charge sources gravity. This is the minimal realization of the idea that charge and mass are two aspects of a unified complex charge.

The factor $(1+\kappa^2)^{-1}$ appears as a global renormalization of both Coulomb and Newtonian forces, slightly weakening them relative to their standard forms. In the limit $\kappa \to 0$, we recover the familiar Coulomb and Newtonian fields separately. For small $\kappa$, the cross-terms represent a tiny but potentially detectable violation of the usual separation between electromagnetic and gravitational phenomena.

\subsection{Force Law and the Violation of the Weak Equivalence Principle}

The equation of motion for a test particle follows from the Lorentz-force analogue in the complex formulation. For a particle with complex charge $\cQ_t = q_t + i\lambda m_t$, the force is:
\begin{equation}
\mathbf{F} = \text{Re}\left[ \cQ_t (\cE + \mathbf{v} \times \cB) \right].
\end{equation}

Consider now the motion in a purely gravitational field, where $\EW = \gNewton$ and $\EV = 0$ (or negligible). The acceleration becomes:
\begin{equation}
\mathbf{a} = -\mathbf{g} + \kappa \frac{q_t}{m_t} \mathbf{g}.
\label{eq:acceleration}
\end{equation}

The first term is the standard gravitational acceleration, independent of the test particle’s composition. The second term, proportional to $\kappa$, depends explicitly on the charge-to-mass ratio $q_t/m_t$ of the test body. This constitutes a \textbf{direct violation of the weak equivalence principle}: bodies with different $q/m$ ratios fall with different accelerations in the same gravitational field.

For two test masses with charge-to-mass ratios $(q/m)_1$ and $(q/m)_2$, the differential acceleration normalized by $g$ is:
\begin{equation}
\boxed{\;
\frac{\Delta a}{g} = \kappa \left[ \left(\frac{q}{m}\right)_1 - \left(\frac{q}{m}\right)_2 \right]
\;}.
\label{eq:equivalence_violation}
\end{equation}

Equation \eqref{eq:equivalence_violation} is the central testable prediction of the model. It is linear in $\kappa$ and linear in the difference of $q/m$ ratios. Crucially, for neutral matter ($q=0$), the violation vanishes, explaining why existing high-precision WEP tests, which deliberately use neutral or nearly neutral test masses, are blind to this effect. The prediction occupies virgin experimental territory precisely because it targets a regime ($q/m$ varied and non-zero) that has been systematically avoided to suppress electrostatic backgrounds.

\subsection*{Phenomenological stance and limitations}
\label{subsec:phenomenological_stance}

This model is presented as a \emph{minimal phenomenological parameterization} that captures a possible charge‑induced violation of the weak equivalence principle. We have deliberately chosen a linear formulation in the complex current to maintain transparency and direct testability, leaving more complete theoretical extensions (variational actions, embedding in non‑linear general relativity) for future work, should experimental evidence warrant it. Our primary aim is to highlight an experimental regime that, to the best of our knowledge, \emph{has not been systematically investigated} in high‑precision WEP tests—since such experiments deliberately suppress electrostatic backgrounds by minimizing charge‑to‑mass ratios—and to provide a clean, one‑parameter target for a controlled laboratory test where $q/m$ is varied intentionally.

\subsection{Contrast with Other Approaches to EM--Gravity Coupling}

The minimalism of our proposal stands out when contrasted with other routes to coupling electromagnetism and gravity. In theories with extra dimensions or dynamical scalar fields \cite{Overduin1997, Burgess2004}, the coupling is typically environment-dependent or mediated by new light particles, leading to composition-dependent but not necessarily charge-linear effects. Axion-like couplings \cite{Wilczek1987}, for instance, produce optical birefringence and CP-violating effects rather than a straightforward $q/m$-dependent violation of the equivalence principle. Non-minimal couplings in curved spacetime (such as $RF_{\mu\nu}F^{\mu\nu}$) generally introduce energy or curvature dependence \cite{Buchbinder2022}, not a clean linear dependence on charge-to-mass ratio. Proposals from supersymmetry or string-inspired scenarios often predict forces that depend on baryon number or spin, but not directly on $q/m$ in the simple linear form derived here \cite{Dimopoulos1996, Polchinski1998}.

What distinguishes the present framework is its \emph{parsimony}: a single dimensionless constant $\kappa$, no new dynamical fields, no extra dimensions, and a prediction that is both sharp and immediately testable with existing laboratory technology. The signature $\Delta a/g \propto q/m$ is unique to this minimal complex-conservation approach, offering a clear target for experimental exploration.

\section{Experimental Pathway and Theoretical Implications}

The linear prediction $\Delta a/g = \kappa(q/m)$ points to a clear experimental signature: a differential acceleration between test bodies with different charge-to-mass ratios. In contrast to traditional equivalence-principle tests that minimize $q/m$ to avoid electrostatic backgrounds \cite{Touboul2017,Wagner2012}, the proposed approach intentionally varies $q/m$ as an independent variable. Torsion balances, with their exceptional torque sensitivity and well-understood techniques for controlling electromagnetic disturbances \cite{Wagner2012,Lamoreaux1997}, offer a natural platform for such a measurement. The key innovation is methodological: instead of suppressing charge, one controls and modulates it, turning a traditional nuisance into the central observable.

Other experimental platforms could complement torsion-balance tests. Free-fall experiments in vacuum towers \cite{Nobili2012}, space-based microgravity missions \cite{Touboul2017}, and atom interferometry with charged particles \cite{Hamilton2015} each offer different systematics and sensitivity profiles. What unites them is the focus on the previously avoided regime where $q/m$ is non-negligible and systematically varied.

The theoretical implications are equally sharp. A non-zero $\kappa$ would indicate that electric charge and gravitational mass are not independent attributes but different components of a unified complex charge, with conservation laws intertwined. This would represent a minimal violation of the traditional separation between electromagnetism and gravity, one that could be embedded in an effective field theory framework without introducing new dynamical fields or extra dimensions \cite{Burgess2004,Donoghue1994}. Conversely, a null result would place a new empirical constraint on linear charge-mass mixing and lend support to a deeper symmetry protecting the distinctness of the two interactions \cite{Will2014}.

Thus, the proposal bridges a long-overlooked gap in experimental gravity and provides a clear phenomenological target. Its minimalism ensures that any outcome, whether a discovery or a new constraint, will inform the broader quest to understand how gravity couples to the other fundamental interactions.

\section{Conclusion}

A simple, yet fundamental question, \emph{Does gravity care about electric charge?}, has led to a model of remarkable economy and an untested experimental frontier. By unifying charge and mass conservation into a single complex current, we have derived the minimal coupling between electromagnetism and linearized gravity, encapsulated in a single dimensionless parameter $\kappa$ and the prediction $\Delta a/g = \kappa(q/m)$.

What is most revealing is not merely the theoretical result, but the methodological blind spot it exposes: precision tests of the equivalence principle, despite their extraordinary sensitivity, have operated in the limit of vanishing charge-to-mass ratio. In avoiding electrostatic backgrounds, they have left a fundamental aspect of gravity unexplored. Our proposal inverts that logic: treat $q/m$ as a controlled variable, not a nuisance, and use a modified torsion balance, or other precision platforms, to directly probe whether gravitational acceleration depends on electric charge.

Whatever the experimental outcome, the implications are profound. A non-zero $\kappa$ would mean that electric charge and gravitational mass are different faces of a unified quantity, with conservation laws intertwined. A null result would instead reinforce a deep symmetry that protects their independence. Either way, we learn something essential about how nature separates, or connects, its fundamental interactions.

This work underscores the power of theoretical minimalism. In a landscape often dominated by elaborate unification schemes, returning to elementary questions and the simplest possible structures can uncover overlooked opportunities. The question we started with is both simple and consequential, and it is now ripe for an experimental answer.

\vspace{0.2cm}
\noindent\textbf{The way forward is clear:} test whether gravity distinguishes between charged and neutral matter. Whatever the answer, it will shed new light on one of the oldest puzzles in physics: how gravity fits into the fabric of the other forces.


\end{document}